\documentclass[aps,twocolumn,showpacs]{revtex4}

\usepackage{epsfig}

\newcommand{\beq}{\begin{equation}}
\newcommand{\eeq}{\end{equation}}
\newcommand{\beqa}{\begin{eqnarray}}
\newcommand{\eeqa}{\end{eqnarray}}
\newcommand{\beqar}{\begin{eqnarray*}}
\newcommand{\eeqar}{\end{eqnarray*}}

\newcommand{\D}{\Delta}
\newcommand{\eps}{\epsilon}

\renewcommand{\l}{\lambda}

\newcommand{\z}{\zeta}

\newcommand{\ie}{{\it i.e.,}\ }
\newcommand{\reef}[1]{(\ref{#1})}
\newcommand\prt{\partial}

\newcommand\cO{{\cal O}}

\newcommand\hl{\hat{\lambda}}

\begin{document}
\tighten
\bibliographystyle{apsrev}
\def\half{{1\over 2}}
\def \D {\mbox{D}}
\def\curl {\mbox{curl}\,}
\def \ep {\varepsilon}
\def \lleq {\lower0.9ex\hbox{ $\buildrel < \over \sim$} ~}
\def \ggeq {\lower0.9ex\hbox{ $\buildrel > \over \sim$} ~}
\def\beq{\begin{equation}}
\def\eeq{\end{equation}}
\def\ber{\begin{eqnarray}}
\def\eer{\end{eqnarray}}
\def \apl {ApJ, }
\def \aps {ApJS, }
\def \pd {Phys. Rev. D, }
\def \prl {Phys. Rev. Lett., }
\def \pl {Phys. Lett., }
\def \np {Nucl. Phys., }
\def \l {\Lambda}

\title{Cosmology from Rolling Massive Scalar Field on the\\
anti-D3 Brane of
de Sitter Vacua} 

\author{Mohammad R.~Garousi}
\affiliation{Department of Physics, Ferdowsi University, P.O.Box 1436, Mashhad, Iran\\
and\\
Institute for Studies in Theoretical Physics and Mathematics IPM\\
P.O.Box 19395-5531, Tehran, Iran} \email{garousi@ipm.ir}
\author{M.~Sami}
\altaffiliation[On leave from:]{ Department of Physics, Jamia
Millia, New Delhi-110025} \email{sami@iucaa.ernet.in}
\affiliation{IUCAA, Post Bag 4, Ganeshkhind,\\
 Pune 411 007, India.}
\author{Shinji Tsujikawa}
\affiliation{Institute of Cosmology and Gravitation, University
of Portsmouth, Mercantile House,
Portsmouth PO1 2EG, \\
United Kingdom} \email{shinji.tsujikawa@port.ac.uk}
\begin{abstract}
We investigate a string-inspired scenario associated with a
rolling massive scalar field on D-branes and  discuss its
cosmological implications. In particular, we discuss cosmological
evolution of the massive scalar field on the ant-D3 brane of KKLT
vacua. Unlike the case of tachyon field, because of the warp
factor of the anti-D3 brane, it is possible to obtain the
required level of the amplitude of density perturbations. We
study the spectra of scalar and tensor perturbations generated
during the rolling scalar inflation and show that our scenario
satisfies the observational constraint  coming from the Cosmic
Microwave Background anisotropies and other observational data.
We also implement the negative cosmological constant arising from
the stabilization of the modulus fields in the KKLT vacua  and
find that this leads to a successful reheating in which the energy
density of the scalar field effectively scales as a pressureless
dust. The present dark energy can be also explained in our
scenario provided that the potential energy of the massive
rolling scalar does not exactly cancel with the amplitude of the
negative cosmological constant at the potential minimum.
\end{abstract}

\maketitle

\section{Introduction}

Cosmological inflation has become an integral part of the standard
model of the universe \cite{review}. Apart from being capable of
removing the shortcomings of the standard big-bang cosmology,
this paradigm has gained a good amount of support from the
accumulated observational data. The recent measurement of the
Wilkinson Microwave Anisotropy Probe (WMAP) \cite{WMAP1,WMAP2} in
the Cosmic Microwave Background (CMB) made it clear that (i) the
current state of the universe is very close to a critical density
and that (ii) primordial density perturbations that seeded
large-scale structure in the universe are nearly scale-invariant
and Gaussian, which are consistent with the inflationary paradigm.

Inflation is often implemented with a single or multiple
scalar-field models \cite{LR}. In most of these models, at least
one of the scalar fields undergoes a slow-roll period allowing an
accelerated expansion of the universe. It then enters the regime
of quasi-periodic oscillations, quickly oscillates and decays
into particles leading to reheating. The late time acceleration
of universe is supported by observations of high redshift
supernovae and indirectly, by observations of the cosmic
microwave background and galaxy clustering. The cosmic
acceleration can be sourced by an exotic form of matter (dark
energy) with a large negative pressure \cite{phiindustry}.
Therefore, the standard model in order to comply with the logical
consistency and observation, should be sandwiched between
inflation at early epoch and quintessence at late times. It is
natural to ask whether one can construct a natural cosmological
model using scalar fields to join the two ends without disturbing
the thermal history of the universe. Attempts have been made to
unify both these concepts using models with a single scalar field
\cite{unifiedmodels}.
\par
Inspite of all the attractive features of cosmological inflation,
its mechanism of realization still remains ad hoc. As inflation
operates around the Planck's scale, the needle of hope points
towards the string theory. It is, therefore, not surprising that
M/String theory inspired models are under active consideration in
cosmology at present. It was recently been suggested that a
rolling tachyon condensate, in a class of string theories, might
have interesting cosmological consequences. Using the boundary
conformal field theory (BCFT) technique, Sen \cite{s1} has shown
that the decay of D-branes produces a pressure-less gas with a
finite energy density that resembles a classical dust. He also
shown that the same results can be extracted from the tachyon DBI
effective action \cite{as1}. Rolling tachyon matter associated
with unstable D-branes has an interesting equation of state
$w$ which smoothly interpolates between $-1$ and 0.
As the tachyon field rolls down the hill, the universe undergoes
an accelerated expansion and at a particular epoch, the scale
factor passes through the point of inflection marking the end of
inflation. At late times the energy  density of tachyon matter
scales as $a^{-3}$, where $a$ is a scale factor. The tachyonic
matter was, therefore, thought to  provide an explanation for
inflation at the early epochs and could contribute to some new
form of cosmological dark matter at late times
\cite{tachyonindustry}. Unfortunately, the effective potentials
for rolling tachyon do not contain free parameters that could be
tuned to make the roll sufficiently slow to obtain enough
inflation and required level of density perturbations \cite{KL}.
The situation could be remedied by invoking the large number of
D-branes separated by distance much larger than $l_s$ (string
scale). However, the number of branes turns out to be typically
of the order of $10^{10}$. This scenario also faces difficulties
associated with reheating and the formation of acoustics/kinks
\cite{Frolov}.\par

In this paper we consider a DBI type effective field theory of
rolling massive scalar boson on the D-brane or anti-D brane
obtained from string theory. We then consider this effective
action for the massive excitation of the anti-D3 brane of the
KKLT vacua, and study the cosmological evolution of the scalar
rolling from some initial value.  The warp factor, $\beta$, of
the anti-D3 brane provides us an interesting possibility to
resolve the problem of the large amplitude of density
perturbations. We also take into account the contribution of the
negative cosmological constant arising from the stabilization of
the modulus fields in the KKLT vacua \cite{KKLT}. This is
important to avoid that the energy density of the rolling massive
scalar over dominates the universe after inflation. The present
critical density can be explained by considering both the minimum
potential of the rolling scalar and the negative cosmological
constant. We also evaluate the inflationary observables such as
the spectral index of scalar perturbations and the
tensor-to-scalar ratio, and examine the validity of this scenario
by using a complication of latest observational data.

\section{Scalar rolling }

Sen has discussed in \cite{s2} a general iterative procedure for
constructing, in string field theory, a one parameter
time-dependent solution describing the rolling of a tachyon away
from its maximum. In the Wick rotated theory, the solution to
order $\l^2$ is
\begin{eqnarray}
|\Psi \rangle &=&\l |\phi_0 \rangle-
\l^2\frac{b_0}{L_0}|\phi_0*\phi_0\rangle+\cO(\l^3)\,, \label{eq00}
\end{eqnarray}
where $L_0$ and $b_0$ are the zero mode of the world-sheet
energy-momentum tensor and the ghost field $b$, respectively. The
higher order terms involve on-shell scattering amplitude of
external states $|\phi_0 \rangle$. This state is related to the
zero momentum tachyon vertex operator $V_T(0)$,\ie
\begin{eqnarray}
|\phi_0\rangle=\cos(\omega X(0))V_T(0)c_1|0 \rangle\,,
\end{eqnarray}
where $c_1$ is the first mode of the ghost field $c$, the
world-sheet field $X$ is the Wick rotation of the space-time time
coordinate $X^0$, and $\omega$, at leading order, is the mass of
the tachyon field. At higher order, $\omega$ is a function of the
parameter $\l$ \cite{s2}. One  can obtain a time-dependent
solution after an inverse Wick rotation $x=-ix^0$ on the final
result.

As has been discussed in \cite{s2}, one can make use of the above
method to generate a one parameter time-dependent solution
describing the oscillation of a positive mass$^2$ scalar field
about the minimum of its potential.   In this case there is no
need for Wick rotation. For a scalar field $\phi$ of mass $m$,
$|\phi_0 \rangle$ is given by
\begin{eqnarray}
|\phi_0 \rangle&=&\cos(\omega X^0(0))V_{\phi}(0)c_1|0 \rangle\,,
\label{eqq1}
\end{eqnarray}
where again $V_{\phi}$ is the zero momentum vertex operator of
the scalar field. At leading order, $\omega=m$ and at higher
order it is a function of the parameter $\l$.

One can also find the BCFT associated with the above massive
scalar rolling solution.  Since the final solution in string
field theory \reef{eq00} is obtained by iterating the initial
solution \reef{eqq1}, one may expect that this corresponds to a
BCFT that is obtained by perturbing the original BCFT by the
boundary term
\begin{eqnarray}
\hl\int dt \cos(\omega X^0(t))V_{\phi}(t)\,, \label{0eqq1}
\end{eqnarray}
where $\hl=\l+\cO(\l^2)$ and $t$ is a parameter labeling the
boundary of the world-sheet. At the leading order, $\omega=m$ and
at higher order it is a function of $\l$. In this case also the
higher-order terms are related to on-shell scattering amplitude
of the  above  vertex operator \cite{s2}. The  solution obtained
in this way describes a one parameter family of BCFT, each member
of which is related to the boundary conformal field theory
describing the original D-brane system by a nearly marginal
deformation. Time evolution of the sources of massless closed
string fields can be extracted from the boundary state
\cite{jp,mbg,sen2}. Unfortunately, the perturbed BCFT is not
solvable, and hence we cannot explicitly compute the boundary
state associated with this BCFT \cite{s2}.

In the present paper, however, we are interested in an effective
action that might produce the above one parameter solution in
field theory. Since the above solution has one parameter, the
effective action should have only first derivative of the scalar
field \cite{comment}. Moreover, we expect that the effective
action should have non-abelian $U(N)$ gauge symmetry when the
original D-brane  system involves  $N$ coincident  D-branes
\cite{witten}, or non-commutative $U(1)$ gauge symmetry when the
original D-brane system carries background B-flux \cite{nsw}.

Recently one of the present authors \cite{mrg1} discussed an
effective action that includes a first derivative of the scalar
fields which have the following vertex operators
\begin{eqnarray}
V_{\phi}&=&\z_i\prt ^n X^ie^{ik\cdot X}\,. \label{eqq2}
\end{eqnarray}
In above relation, the index $i$ runs over the transverse
directions of the D$_p$-brane, \ie $i=p+1,\cdots, 25$, $\z_i$
represent the polarization of the scalar state, and $k^a$ with
$a=0,1,\cdots, p$ is the momentum of the scalar state along the
D$_p$-brane. Mass of the above vertex operator is
$m^2=(n-1)/{\alpha'}$ where $n$ is an integer number and
$\sqrt{\alpha'}$ is the string length scale. Hence, for $n\geq 2$
this vertex represents a massive scalar state. The disk level
four-point amplitude of the above scalar has been evaluated in
Ref.\,\cite{mrg1}. Then an expansion for the amplitude has been
found that its leading order terms correspond to an action with a
non-abelian gauge symmetry. Reducing the non-abelian symmetry to
the abelian one in which we are interested, the leading couplings
are consistent with the following Born-Infeld type action
\cite{mrg1}:
\begin{eqnarray}
S&=&-\int d^{p+1}x V(\phi) \label{action} \\
& & \times \sqrt{-\det(\eta_{ab}+2\pi\alpha'
F_{ab}+\prt_a\phi^i\prt_b\phi_i)}\nonumber\,,
\end{eqnarray}
where $F_{ab}$ and $\phi^i$ are the gauge field strength and the
scalar fields, respectively. The massive scalar potential is
\begin{eqnarray}
V(\phi)&=&T_p\left(1+\frac{1}{2}m^2\phi^i\phi_i+\frac{1}{8}m^4
(\phi^i\phi_i)^2+\cdots\right) \nonumber
\\
&=&T_pe^{\frac{1}{2}m^2\phi^i\phi_i}\,, \label{eq2}
\end{eqnarray}
where $T_p$ is the D$p$-brane tension. In the second line above
we have speculated a closed form for the expansion.

The massless closed string fields can be added into the above
effective action by the general grounds of covariance, T-duality,
and by the fact that the world-sheet is disk, that is
\begin{eqnarray}
S&=&-\int d^{p+1}x V(\phi)e^{-\Phi} \label{eeq3} \\
& & \times \sqrt{-\det(g_{ab}+B_{ab}+2\pi\alpha' F_{ab}
+\prt_a\phi^i\prt_b\phi_i)}, \nonumber
\end{eqnarray}
where $\Phi$, $g_{ab}$ and $B_{ab}$ are the dilaton, metric and
the anti-symmetric two tensor fields, respectively.

One may  use the world-sheet conformal field theory technique
\cite{mrgrm} to evaluate the S-matrix element of two massive
scalar states \reef{eqq2} and one massless closed string vertex
operators to confirm the closed string couplings in the above
action. The scalar fields are not massless so, unlike the
massless case, the expansion of the amplitude will not be a
low-energy expansion. In order to find an appropriate expansion
for the S-matrix element, one may firstly evaluate the amplitude
in  the presence of the background B-flux \cite{mrg2}. The
amplitude has then massless pole and infinite tower of massive
poles. Using the fact that the effective field theory in this
case is a non-commutative field theory, and that the
non-commutative massive field theory has graviton-gauge field
coupling as well as the scalar-scalar-gauge field coupling, then,
one finds that the expansion of the amplitude should be  around
the massless pole. After finding the expansion, one may set the
background B-flux to zero. The leading non-zero term of the
expansion should then  be consistent with the coupling of the
scalar-scalar-massless closed string field extracted from the
above action.

The construction of the massive effective action from S-matrix
element can easily be carried out in the superstring theory. One
needs only consider the analog of the  massive vertex operators
\reef{eqq2} in the superstring theory. It is argued in
\cite{mrg1} that the action \reef{action} is consistent with the
leading order terms of the S-matrix element of four massive
vertex operators in the superstring theory. The index $i$ in this
case takes the values $i=p+1, \cdots, 9$, and
$m^2=(n-1)/(2\alpha')$ with $n\geq 1$ for BPS D-branes in which we
are interested. On the general grounds, one expects that the
closed string fields $\Phi$, $B_{ab}$ and $g_{ab}$ have the
coupling consistent with the action \reef{eeq3}. In the
superstring, there is also the RR massless closed string fields.
One may study the S-matrix element of two massive scalar and one
RR vertex operators to find the coupling of RR to the scalar
fields. However, we are not interested in these couplings here.

In the present  paper, we are interested in the cosmological
evolution of the massive scalar rolling of BPS-D3-brane of type
IIB string theory. In principle one can study this evolution in
string field theory or in the BCFT as we already mentioned above.
However, no analytical solution can be obtained in string field
theory, and BCFT is not solvable in this case \cite{s2}. Hence we
stick to the effective action \reef{eeq3} and find the
cosmological evolution in this field theory.

In order to study the cosmological evolution of the D$3$-brane or
anti-D$3$ brane, one has to assume that the extra six dimensions of
type IIB string theory are frozen in a compact manifold such that
evolution of the massive scalar field on the brane does not
decompactify  the internal manifold . Recently a construction of
this compactification was reported by the authors
in Ref.~\cite{KKLT} (KKLT). In the next
section we review this construction.

\section{Review of KKLT vacua}

The low energy effective action of string/M-theory in four
dimension is described by $N=1$ supergravity,
\beqa
S&=&\int
d^4x\,\sqrt{-g}\biggl[\frac{M_p^2}{2}R+
g^{\mu\nu}K_{a\bar{b}}\partial_{\mu}
\varphi^a\partial_{\nu}\bar{\varphi}^b
\nonumber\\
&&-e^{K/M_p^2}\left(K^{a\bar{b}}D_a
WD_{\bar{b}}\bar{W}-\frac{3}{M_p^2}|W|^2\right)
\nonumber \\
& &+\cdots \biggr],
\eeqa
where $a,b$ run over all complex moduli fields. In the above
equation, the holomorphic function $W(\varphi^a)$ is the
superpotential, $K(\varphi^a,\bar{\varphi}^b)$ is the Kahler
potential,
$K_{a\bar{b}}=\partial^2K/\partial\varphi^a\partial\bar{\varphi}^b$
is the Kahler metric, and $D_aW$ is the Kahler derivative,
\beqa
D_aW&=&\frac{\partial
W}{\partial\varphi^a}+\frac{W}{M_p^2}\frac{\partial
K}{\partial\varphi^a}\,.
\eeqa The supersymmetry will be unbroken only
for the vacua in which $D_aW=0$ for all $a$; the effective
cosmological constant is thus non-positive. Some preferable
choices of the Kahler potential $K$, and superpotential $W$ will
be selected at the level of the fundamental string/M-theory. We
set $M_p^2=1$ in the rest of this section.

Using the flux compactification  of Type IIB string theory
\cite{pm,sg}, the authors in Ref.~\cite{sg}
used the following tree level
functions for $K$ and $W$:
\beqa
K&=&-3\ln[-i(\rho-\bar{\rho})]-\ln[-i(\tau-\bar{\tau})]
\nonumber \\
&&-\ln[-i\int_M\Omega\wedge\bar{\Omega}]\,, \\
W&=&\int_MG_3\wedge\Omega\,,
\eeqa
where $\Omega$ is the holomorphic
three-form on the Calabi-Yau space and $G_3=F_3-\tau H_3$ where
$F_3$ and $H_3$ are the R-R flux and the  NS-NS flux,
respectively, on the 3-cycles of the internal Calabi-Yau
manifold, $\rho$ the volume modulus which includes the volume of
Calabi-Yau space and an axion field coming from the R-R 4-form,
$C_{(4)}$,  and $\tau=C_{(0)}+ie^{-\Phi}$ is axion-dilaton
modulus. Since $W$ is not a function of $\rho$,one has
$K^{\rho\bar{\rho}}D_{\rho}WD_{\bar{\rho}}\bar{W}=3|W|^2$,
which reduces the supergravity potential to
\beqa {\cal V}&=&
e^{K}\left(K^{i\bar{j}}D_iWD_{\bar{j}}\bar{W}\right)\,,
\eeqa where
$i,j$ run over all moduli fields except $\rho$.

It is argued in \cite{sg} that  the condition $D_iW=0$ fixes all
complex moduli except $\rho$. This gives zero  effective
cosmological constant. On the other hand, the supersymmetric
vacua that satisfies $D_{\rho}W=0$ gives $W=0$, whereas,
non-supersymmetric vacua yield $W=W_0\ne 0$.

The flux $F_3$ and $H_3$ are also the sources for a warp factor
\cite{pm,sg}. Therefore, models with flux are generically warped
compactifications,
\beqa
ds^2_{10}&=&e^{2A(y)}g_{\mu\nu}(x)dx^{\mu}dx^{\nu} \nonumber \\
& &+e^{-2A(y)}\tilde{g}_{mn}(y)dy^mdy^n\,,
\label{warp}
\eeqa where the warp
factor, $e^A$, can be computed in the regions closed to a conifold
singularity of the Calabi-Yau manifold \cite{sg}. The result for
the warp factor is exponentially suppressed at the throat's tip,
depending on the fluxes as:
\beqa e^{A_{\rm min}}\sim
\exp\left[-\frac{2\pi N}{3g_sM}\right]\,,
\label{min}
\eeqa
where $g_s$ is the
string coupling constant, and integers $M$, $N$ are the R-R and NS-NS
three-form flux, respectively. While the warp factor is of order
one at generic points in the $y$-space, its minimum value can be
extremely small given a suitable choice of fluxes.

In order to fix $\rho$ as well, KKLT \cite{KKLT} added a
non-perturbative correction \cite{ew} to
the superpotential, that is
\beqa
& &K=K_0-3\ln[-i(\rho-\bar{\rho})]\,, \\
& &W=W_0+Ae^{ia\rho}\,,
\eeqa
where $A$ and $a$ are two constants. In this equation $W_0=\int
G_3\wedge\Omega$ and
$K_0=-\ln[-i(\tau-\bar{\tau})]-\ln[-i\int\Omega\wedge\bar{\Omega}]$
are evaluated at the above fixed moduli. Now the conditions
$D_iW=0$ is automatically  satisfied, and the supersymmetric
condition $D_{\rho}W=0$ gives
$W_0=-Ae^{ia\rho}(1+ia(\rho-\bar{\rho})/3)$ which  fixes $\rho$
in terms of $W_0$. They also produce a negative cosmological
constant, that is
\beqa
{\cal V}_{AdS}=-3e^{K}|W|^2\,,
\eeqa where $K$
and $W$ in above equation should be evaluated at the fixed moduli
including $\rho$. Therefore, all the moduli are stabilized while
preserving supersymmetry with a negative cosmological constant.

In order to obtain a de Sitter (dS) vacuum, KKLT introduced anti-D3
brane, and in so doing break the supersymmetry of the above
Anti de Sitter (AdS)
vacuum. The background fluxes generate a potential for the
world-volume scalars of  the anti-D3 brane, hence, it does not
introduce additional moduli \cite{sk}. The anti-D3 brane, however,
adds an additional energy to the supergravity potential
\cite{sk}:
\beqa
\delta {\cal
V}&=&\frac{2a_0^4T_3}{g_s^4}\frac{1}{({\rm Im}\rho)^3}\,,
\eeqa
with $a_0$ the warp factor at the location of the anti-D3 brane, and
$T_3$ the brane tension. Because of the warping the anti-D3 brane
energetically prefers to sit at the throat's tip that has a minimum
warped factor.  By tuning the fluxes which inter in Eq.~(\ref{min}),
one can perturb the above AdS vacua to produce dS vacua with a
tunable cosmological constant, that is
\beqa {\cal V}_{\rm
dS}=\beta^2T_3-\Lambda\label{vds}\,,
\eeqa
where
$\beta^2=2a_0^4/[g_s^4({\rm Im}\rho_{cr})^3]$ and $\Lambda$ is
the negative cosmological constant of the AdS vacua. The
effective action with all moduli stabilized is then
\beqa S=\int
d^4x\left\{\sqrt{-g}\left(\frac{R}{2\kappa^2}+\Lambda\right)
-\beta^2T_3\sqrt{-g}\right\}.
\eeqa
By adding  a string-inspired scalar field (inflaton)
to this action, one can study various cosmological model in
string theory.

\section{The Cosmological Model}

Most of the cosmological model in the KKLT vacua considers another
mobile D3 brane in the compact space \cite{KKLMMT}. In this
setting the distance moduli between D3 brane and anti-D3 brane
plays the rule of inflaton. The cosmological scenario in this
setting should be the following: When the mobile D3-brane is far
from the constant anti-D3 brane the motion of D3-brane gives rise
to inflation. When the brane reaches to a critical distance from the
anti-D3 brane the scalar field converts to a tachyonic mode which causes
brane-anti-brane annihilation. This process makes a naturally
graceful exit from inflation and is expected to produce radiation
\cite{jmc}.

Adding the mobile D3 brane to the KKLT vacua introduces  some new
moduli and ruins the nice feature of the all moduli stabilized
KKLT vacua \cite{bsa,KKLT}.  
It is shown  in \cite{KKLMMT} as  how to stabilize all moduli in this case,
however,  volume stabilization modifies the inflaton potential and renders
it too steep for inflation.

Our  cosmological model  does not introduce any new moduli to the
KKLT vacua. It, instead,  considers a massive open string
excitation of the anti-D3 brane as the inflaton. The cosmological
scenario should be the following. Rolling of this scalar field,
$\phi$, from an initial value $\phi_0$ towards the minimum of its
potential generates inflation when $\phi$ is far from its ground
state $\phi=0$, reheats the universe  when $\phi$ oscillates
around its minimum at $\phi=0$, and eventually mimics  the KKLT
cosmological constant (\ref{vds}) when it sits at $\phi=0$\cite{Linde,becker}. 

The initial value of the scalar field in the inflation epoch is
far from its ground state, hence, one should use an effective
action for the scalar field which includes all power of $\phi$.
We use the DBI type  action introduced before  as the effective
action  for the massive scalar field $\phi$. When the anti-D3 brane
is in a generic point in the compact space, the scalar and metric
fields has the action
\beqa
S=-\int d^4x\, V(\phi)\sqrt{-\det(g_{ab}+
\prt_{a}\phi\prt_{b}\phi)}\,,
\label{total}
\eeqa
where the scalar
field $\phi$ has dimension  $\sqrt{\alpha'}$ and $V(\phi)$ is
given by
\begin{equation}
V(\phi)=T_3 e^{{1 \over 2}m^2 \phi^2}\,. \label{pot3}
\end{equation}
However, the anti-D3 brane in the KKLT vacua is in the warp
metric \reef{warp} with warp factor $\beta$. The action
\reef{total} for this metric becomes
\begin{eqnarray}
S&=& -\int d^4x \beta^2V(\phi)\sqrt{-\det(g_{ab}+
\beta^{-1}\prt_a\phi\prt_b\phi)}.
\nonumber\\
\label{eq01}
\end{eqnarray}
Normalizing the scalar field as $\phi\rightarrow
\sqrt{\beta}\phi$, one finds the standard Born-Infeld
type action
\begin{eqnarray}
S=-\int d^{4}x V(\phi)\sqrt{-\det(g_{ab}+\prt_{a}
\phi\prt_{b}\phi)}\,, \label{eq02}
\end{eqnarray}
where now the potential is
\begin{eqnarray}
V(\phi)=V_0e^{\frac{1}{2}m^2 \beta \phi^2}\,,~~ {\rm
with}~~V_0=\beta^2 T_3\,. \label{eq03}
\end{eqnarray}
The constant $V_0$ can be less than $T_3$ for small values of
$\beta$ with $\beta<1$. Considering this effective action for the
massive open string excitation of the anti-D3 brane of the KKLT
vacua, one finds the following effective action
\beqa
S&=& \int
d^4x\biggl\{\sqrt{-g}\left(\frac{R}{2\kappa^2}+\Lambda\right)
\nonumber \\
& &-V(\phi)\sqrt{-\det(g_{ab}+
\partial_{a}\phi\partial_{b}\phi)}\biggr\}\,.
\label{eq0}
\eeqa
We will consider cosmological evolution of the massive scalar
rolling using the above effective action.


\section{Inflation from scalar rolling} \label{intro}

In this section we study inflation from the massive scalar field
rolling on the anti-$D3$ brane.  In a spatially flat
Friedmann-Robertson-Walker (FRW) background with a scale factor
$a$, the energy momentum tensor for the Born-Infeld scalar $\phi$
acquires the diagonal form $T^{\mu}_{\nu}={\rm
diag}\left(-\rho,p,p,p\right)$, where the energy density $\rho$
and the pressure $p$ are given by [we use the signature $(-, +, +,
+)$],
\begin{eqnarray}
\rho &=& {V(\phi) \over {\sqrt{1-\dot{\phi}^2}}}-\Lambda\,,\\
p &=& -V(\phi)\sqrt{1-\dot{\phi}^2}+\Lambda\,,
\end{eqnarray}
where a dot denotes a derivative with respect
to a cosmic time, $t$.

The Hubble rate, $H \equiv \dot{a}/a$, satisfies the Friedmann
equation
\begin{eqnarray}
H^2={1 \over 3M_p^2} \rho = {1 \over 3M_p^2}\left({V(\phi) \over
{\sqrt{1-\dot{\phi}^2}}}-\Lambda\right)\,. \label{H}
\end{eqnarray}
The equation of motion of the rolling scalar field which follows
from Eq.~(\ref{eq0}) is
\begin{eqnarray}
{\ddot{\phi} \over {1-\dot{\phi}^2}}+3H \dot{\phi}+{V_{,\phi}
\over V({\phi})}=0\,. \label{ddphi}
\end{eqnarray}
The conservation equation equivalent to Eq.~(\ref{ddphi}) has the
usual form
\begin{eqnarray}
\dot{\rho}+3H (1+w)\rho=0\,,
\end{eqnarray}
where the equation of state for the field $\phi$ is
\begin{eqnarray}
w \equiv p/\rho=
-1+\frac{\dot{\phi}^2}{1-c\sqrt{1-\dot{\phi}^2}}\,,
\label{eqst}
\end{eqnarray}
with $c \equiv \Lambda/V(\phi)\leq1$.
The conservation equation formally integrates to
\begin{equation}
\rho=\rho_0 e^{-3 \int{{da \over a}(1+w)}} \,,
\label{scaling}
\end{equation}
where $\rho_0$ is a constant.
In the inflation epoch we have $\beta
m^2\phi\gg1$, which gives $c\ll 1$. This clearly demonstrates that
the field energy density can not scale faster than $1/a^3$ ($-1
\le w < 0$) during this epoch inspite of the steepness of the field
potential. Obviously, this is inbuilt in the evolution equation
(\ref{ddphi}). However, as we shall show later, in the reheating
epoch the situation changes drastically.
\par The slow-roll parameter for the Born-Infeld scalar
is given by
\begin{eqnarray}
\epsilon \equiv
-\frac{\dot{H}}{H^2} &=& \frac32\left(\frac{\dot{\phi}^2}
{1-c\sqrt{1-\dot{\phi}^2}}\right) \nonumber \\
&\simeq& \frac32\dot{\phi}^2\simeq
\frac{M_p^2}{2}\frac{V{}_{,\phi}^2}{V^3} \,.
\label{eq1}
\end{eqnarray}
In deriving this relation we used the slow-roll approximations,
$\dot{\phi}^2 \ll 1$ and $|\ddot{\phi}| \ll 3H|\dot{\phi}|$ in
Eqs.~(\ref{H}) and (\ref{ddphi}), and also the fact that in
inflation epoch $c\ll 1$. The condition for inflation is
characterized by $\epsilon<1$, which translates into a condition,
$\dot{\phi}^2\ < 2/3$.

With the potential (\ref{eq03}), the slow-roll parameter is
written as
\begin{eqnarray}
\epsilon\simeq\frac{m^4M_p^2}{2T_3}\phi^2 e^{-\frac{1}{2}m^2 \beta
\phi^2}\,. \label{ep}
\end{eqnarray}
The slow-roll condition, $\eps < 1$, can be satisfied for large
$\phi$ due to the presence of the exponential factor. Hence
unlike the tachyon inflation in which inflation happens only
around the top of the potential, it is possible to obtain a
sufficient number of $e$-foldings. Nevertheless it is important
to investigate observational constraints on our scenario, since
this type of inflation typically generates density perturbations
whose amplitudes are too high to match with observations
\cite{KL}. In the next section, we will analyze whether our
scenario agrees with observations of the temperature anisotropies
in Cosmic Microwave Background (CMB).

\section{Density perturbations generated in inflation
due to rolling scalar and observational constraints}

In this section we shall study the spectra of scalar and tensor
perturbations generated in rolling scalar field inflation and
analyze whether our scenario satisfies observational constraints
coming from CMB anisotropies. Hwang and Noh \cite{Hwang} provided
the formalism to evaluate the perturbation spectra for the
general action
\beqa S=\int {\rm d}^4 x\sqrt{-g}\frac12 f(R, \phi, X)\,, \eeqa
which includes our action (\ref{eq0}). Here the function $f$
depends upon the Ricci scalar $R$, a scalar field $\phi$ and its
derivative $X=(\nabla \phi)^2/2$.
The Born-Infeld scalar field corresponds to the case with
\beqa f=\frac{R}{\kappa^2}+2\Lambda-2V(\phi)\sqrt{1+2X}\,.
\label{tach} \eeqa
In this section we use the unit $\kappa^2=1$.

Let us consider a general perturbed metric about the flat FRW
background \cite{MFB}:
\begin{eqnarray}
ds^2 &=& -(1+2A)dt^2 + 2a(t)B_{,i} dx^idt  \nonumber\\
&+& a^2(t)[(1+2\psi) \delta_{ij}+2E_{,i,j}+2h_{ij}] dx^i dx^j.
\nonumber \\
\label{pmetric}
\end{eqnarray}
Here $A$, $B$, $\psi$, and $E$ correspond to the scalar-type
metric perturbations, whereas $h_{ij}$ characterizes the
transverse-traceless tensor-type perturbation. It is convenient
to introduce comoving curvature perturbations, ${\cal R}$,
defined by
\begin{eqnarray}
{\cal R} \equiv \psi-\frac{H}{\dot{\phi}}\delta \phi\,,
\label{calR}
\end{eqnarray}
where $\delta \phi$ is the perturbation of the field $\phi$.

Making a Fourier transformation, one gets the equation of motion
for ${\cal R}$ from the Lagrangian (\ref{tach}), as \cite{Hwang}
\begin{eqnarray}
\frac{1}{a^3Q_{\rm S}} \left(a^3 Q_{\rm S} \dot{\cal
R}\right)^{\bullet} + c_{\rm s}^2\frac{k^2}{a^2}{\cal R}=0\,,
\label{Req}
\end{eqnarray}
with
\begin{eqnarray}
Q_{\rm S} &\equiv& \frac{f_{,X}X+2f_{,XX}X^2}{H^2}\,,\\
c_{\rm s}^2 &\equiv& \frac{f_{,X}X} {f_{,X}X+2f_{,XX}X^2}\,,
\label{Qcs}
\end{eqnarray}
where $k$ is a comoving wavenumber. In our case we have $c_{\rm
s}^2=1-\dot{\phi}^2>0$, which means that there is no instability
for perturbations \cite{comment2}. We shall introduce three
parameters, defined by
\begin{eqnarray}
\epsilon_1 \equiv \frac{\dot{H}}{H^2}\,,~~~ \epsilon_2 \equiv
\frac{\ddot{\phi}}{H\dot{\phi}}\,,~~~ \epsilon_3 \equiv
\frac{\dot{Z}_{\rm S}}{2HZ_{\rm S}}\,, \label{slow}
\end{eqnarray}
where $Z_{\rm S}\equiv -(f_{,X}/2+f_{,XX}X)=
V(1-\dot{\phi}^2)^{-3/2}$. Under the slow-roll approximation,
$|\epsilon_i| \ll 1$, the power spectrum of curvature
perturbations is estimated to be \cite{Hwang}
\begin{eqnarray}
{\cal P}_{\rm S}=\left(\frac{H^2}{2\pi\dot{\phi}} \right)^2
\frac{1}{Z_{\rm S}}\,, \label{powersca}
\end{eqnarray}
together with the spectral index
\begin{eqnarray}
n_{\rm S}-1=2(2\epsilon_1-\epsilon_2-\epsilon_3)\,.
\label{indexsca}
\end{eqnarray}

The tensor perturbation satisfies the equation
\begin{eqnarray}
\ddot{h}_i^j+3H\dot{h}_i^j+\frac{k^2}{a^2} h_i^j=0\,,
\label{tenequ}
\end{eqnarray}
and its spectrum is simply given by
\begin{eqnarray}
{\cal P}_{\rm T}=8\left(\frac{H}{2\pi}\right)^2\,,
\label{powerten}
\end{eqnarray}
together with the spectral index
\begin{eqnarray}
n_{\rm T}=2\epsilon_1\,. \label{indexten}
\end{eqnarray}
Then the tensor-to-scalar ratio is defined as
\begin{eqnarray}
R \equiv \frac{{\cal P}_{\rm T}}{{\cal P}_{\rm S}}
=8\frac{\dot{\phi}^2}{H^2}Z_{\rm S}\,. \label{ratio}
\end{eqnarray}
Note that our definition of $R$ coincides with the one in
Ref.~\cite{cmb1} but differs from Refs.~\cite{cmb2}.

\subsection{The amplitude of scalar perturbations}

Making use of the slow-roll approximation in Eqs.~(\ref{H}) and
(\ref{ddphi}), the amplitude of scalar perturbations is estimated
as
\begin{eqnarray}
P_{\rm S} \simeq \frac{1}{12\pi^2}
\left(\frac{V^2}{V_{,\phi}}\right)^2\,. \label{Psapprox}
\end{eqnarray}
For the potential (\ref{eq03}), one obtains
\begin{eqnarray}
P_{\rm S} \simeq \frac{\beta^2m^4\phi^2} {48\pi^2 \epsilon^2}\,.
\label{Psapprox2}
\end{eqnarray}

The slow-roll parameter $\epsilon$ is smaller than of order unity
on cosmologically relevant scales observed in the COBE satellite.
Then we have $P_{\rm S} \gtrsim \beta^2m^4\phi^2$ for $\epsilon
\lesssim 0.1$. Since $m$ is of order 1 in $1/\sqrt{\alpha'}$
unit, one obtains the relation $P_{\rm S} \gtrsim \beta^2\phi^2$.
In order for the slow-roll parameter to be smaller than 1 during
inflation, we require the condition $\beta \phi^2 \gtrsim 1$ in
Eq.~(\ref{ep}) [note that we are considering the case with $T_3
\sim 1$]. Therefore this gives $P_{\rm S} \gtrsim \beta$, which
means that the observed amplitude, $P_{\rm S} \simeq 10^{-9}$,
can not be explained for $\beta=1$. This is actually what was
criticized in Ref.~\cite{KL} in the context of tachyon inflation.
However, we can avoid this problem by introducing the D-brane in
a warped metric with $\beta$ satisfying $\beta \ll 1$.

Lets us consider other types of the potential to check the
generality of our scenario. In the case of the polynomial
potential, $V(\phi)=T_3\phi^p$, one can make a similar process of
the transformation of variables carried out in from
Eqs.~(\ref{eq01}) to (\ref{eq03}), yielding
\begin{eqnarray}
V(\phi)=V_0\phi^p, \label{potp}
\end{eqnarray}
with $V_0=\beta^{2+p/2}T_3$. In this case the number of
$e$-foldings is
\begin{eqnarray}
N=\int_{t}^{t_f} H {\rm d}t \simeq \frac{V_0}{p(p+2)}
\left(\phi^{p+2}- \phi_f^{p+2}\right)\,, \label{efold}
\end{eqnarray}
where $\phi_f$ is the value of $\phi$ at the end of inflation.
Expressing $\phi$ in terms of $N$, one gets the amplitude
\begin{eqnarray}
P_{\rm S}=\frac{\left[V_0^2
\left\{p(p+2)N\right\}^{2(p+1)}\right]^{\frac{1}{p+2}}} {12\pi^2
p^2}, \label{ampoly}
\end{eqnarray}
where we neglected the second term in the square bracket of
Eq.~(\ref{efold}). In the quadratic potential ($p=2$) we obtain
$P_{\rm S}=\beta^{3/2} \sqrt{2T_3}N^{3/2}/3\pi^2$. Therefore one
can get $P_{\rm S} \sim10^{-9}$ provided that $\beta \sim
10^{-8}$ with $T_3=1$ and $N=55$. Note that it is impossible to
get a right level of the size of density perturbations for
$\beta=1$.

When $p \gg 1$ the amplitude (\ref{ampoly}) is simplified as
\begin{eqnarray}
P_{\rm S}=\frac{\beta p^2 N^2}{12\pi^2}.
\label{amploy2}
\end{eqnarray}
This explicitly shows that $P_{\rm S} \gg 1$ for $\beta=1$.
Inclusion of the $\beta$ term that is much smaller than 1
suppresses the amplitude, which makes it possible to obtain
$P_{\rm S} \sim 10^{-9}$.

We shall also consider an exponential potential
\begin{eqnarray}
V=V_0e^{\lambda \sqrt{\beta}\phi}, \label{expo}
\end{eqnarray}
where $\lambda$ is a constant and $V_0=\beta^2 T_3$. Since the
number of $e$-foldings is estimated as $N=V_0e^{\lambda
\sqrt{\beta}\phi}/(\lambda^2 \beta)$, the amplitude of scalar
perturbations is
\begin{eqnarray}
P_{\rm S} \simeq \frac{\beta \lambda^2N^2}{12\pi^2}\,,
\label{ampexp}
\end{eqnarray}
which takes the similar form to Eq.~(\ref{amploy2}), since the
exponential potential is viewed as the case of $p \gg 1$. With
the choices $\lambda=1$ and $N=55$, one has $P_{\rm S} \sim
10^{-9}$ for $\beta \sim 10^{-9}$.

{}From the above argument, we conclude that the picture of the
D-brane in a warped metric is crucially important to get the
right level of the amplitude of scalar perturbations.

\subsection{Observational constraints \\
in terms of the spectral index $n_{\rm S}$ and the
tensor-to-scalar ratio $R$}

Even if our scenario can satisfy the condition of the COBE
normalization, it is not obvious whether the model can be allowed
from other observational constraints. In the Randall-Sundrum II
braneworld scenario, the steep inflation driven by an exponential
potential is outside of the two-dimensional observational contour
bound in terms of $n_{\rm S}$ and $R$ \cite{cmbbrane}. In this
subsection, we shall investigate whether our scenario lies in
posterior contour bounds in the $n_{\rm S}$-$R$ plane.

Using the slow-roll analysis, we have
\begin{eqnarray}
\epsilon_1 \simeq -\frac{V_{,\phi}^2}{2V^3},~ \epsilon_2 \simeq
\frac32 \frac{V_{,\phi}^2}{V^3}-\frac{V_{,\phi \phi}}{V^2},~
\epsilon_3 \simeq -\frac{V_{,\phi}^2}{2V^3}, \label{slow2}
\end{eqnarray}
and
\begin{eqnarray}
n_{\rm S}-1 &=& -4\frac{V_{,\phi}^2}{V^3} +2\frac{V_{,\phi
\phi}}{V^2},
\\
n_{\rm T} &=& -\frac{V_{,\phi}^2}{V^3},
\\
R &=& 8\frac{V_{,\phi}^2}{V^3}\,. \label{indexsca2}
\end{eqnarray}
This means that the same consistency relation, $R=-8n_{\rm T}$,
holds as in the Einstein gravity, as was pointed out in
Ref.~\cite{SV}
Therefore the observational contour plot derived
using this relation can be applied in our case
as well.
Note that this property holds even in generalized
Einstein theories including the 4-dimensional dilaton gravity and
scalar tensor theories \cite{Burin}.

\begin{figure}
\epsfxsize = 3.0in \epsffile{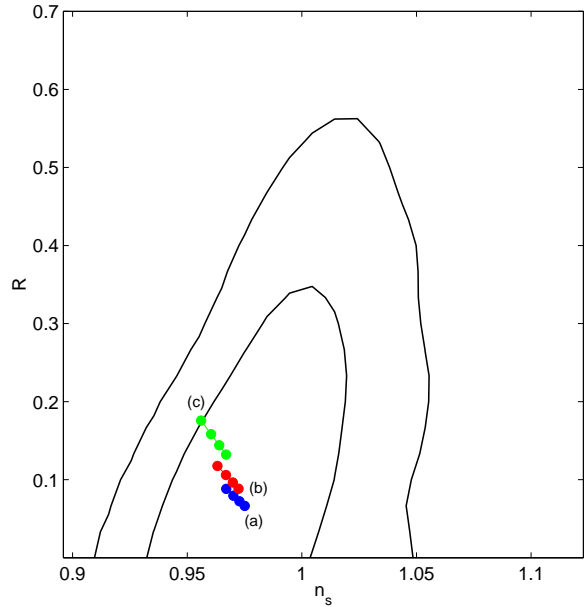} \caption{2D posterior
constraints in the $n_{\rm S}$-$R$ plane with the $1\sigma$ and
$2\sigma$ contour bounds. Each case corresponds to (a) $p=2$, (b)
$p=4$ and (c) $p \gg 1$ for the potential (\ref{potp}) with
$e$-foldings $N=45, 50, 55, 60$ (from top to bottom). The scalar
rolling potential (\ref{eq03}) belongs to the case (c). }
\label{nsandR}
\end{figure}

In the case of the polynomial potential given in (\ref{potp})
$n_{\rm S}$ and $R$ are given as
\begin{eqnarray}
n_{\rm S}-1 =-\frac{2p(p+1)}{V_0 \phi^{p+2}}\,,~~~
R=\frac{8p^2}{V_0\phi^{p+2}}\,. \label{index2}
\end{eqnarray}
Since the value of $\phi$ at the end of inflation can be
estimated as $\phi_f^{p+2}=p^2/(2V_0)$ by setting $\epsilon=1$,
the number of $e$-folds is written as
\begin{eqnarray}
N=\frac{V_0}{p(p+2)}\phi^{p+2}- \frac{p}{2(p+2)}, \label{efold2}
\end{eqnarray}
Combining Eqs.~(\ref{index2}) and (\ref{efold2}), we obtain the
relation
\begin{eqnarray}
n_{\rm S}-1 &=& -\frac{4(p+1)}{p(2N+1)+4N}\,, \\
R &=& \frac{16p}{p(2N+1)+4N}\,. \label{index3}
\end{eqnarray}

For the quadratic potential ($p=2$) one has $n_{\rm S}=1-6/(4N+1)$
and $R=16/(4N+1)$, which yields $n_{\rm S}=0.973$ and $R=0.072$
for a cosmologically relevant scale, $N \simeq 55$. We have
$n_{\rm S}=1-5/(3N+1)$ and $R=16/(3N+1)$ for the quartic
potential ($p=4$), giving $n_{\rm S}=0.970$ and $R=0.096$ for
$N=55$. In both cases the theoretical points are inside the
$1\sigma$ observational contour bound (see Fig.~\ref{nsandR}).

If we take the limit $p  \gg 1$, we can get $n_{\rm S}$ and $R$
corresponding to the exponential potential (\ref{expo}) and our
potential (\ref{eq03}), as
\begin{eqnarray}
n_{\rm S}-1 =-\frac{4}{2N+1}\,, ~~~R=\frac{16}{2N+1}\,.
\label{indextac}
\end{eqnarray}
Actually this result completely coincides with what was obtained
in Ref.~\cite{SV} for an exponential potential. In
Fig.~\ref{nsandR} we plot the theoretical values of $n_{\rm S}$
and $R$ with $e$-folds $N=45, 50, 55, 60$ in 2D posterior
observational constraints. We find that the $e$-folds with $N
\gtrsim 50$ is inside the $1\sigma$ contour bound. Unlike the
Randall-Sundrum II braneworld scenario, the steep inflation
driven by an exponential potential (and even the steeper
potential) is not ruled out in the context of the rolling scalar
field inflation. Therefore the slow-roll inflation driven by the
massive Born-Infeld scalar field satisfies the observational
requirement coming from CMB even when the potential is steep.
\begin{figure}
\resizebox{3.0in}{!}{\includegraphics{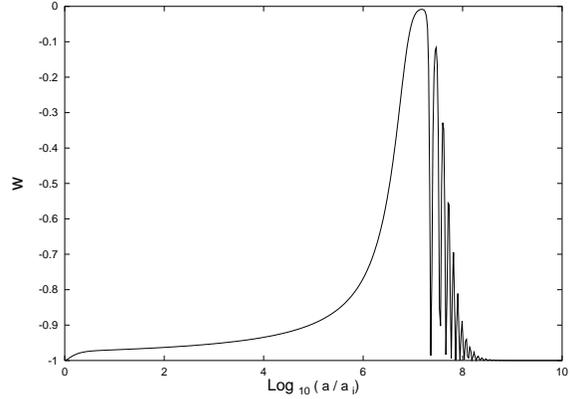}}
\caption{The
evolution of the equation of state parameter $w$ (for massive Born-Infeld scalar field) is shown versus
the scale factor
for a fixed value of the warp factor $\beta=0.01$ in case of $\Lambda=0$. As the
field evolves towards the origin, the parameter $w$ moves towards
zero but never attains it. After this stage, $w$ fast drops and
begins to oscillate such that the average equation of state is
sufficiently negative and finally settles at $-1$.}

\label{bsw}
\end{figure}

\begin{figure}
\resizebox{3.0in}{!}{\includegraphics{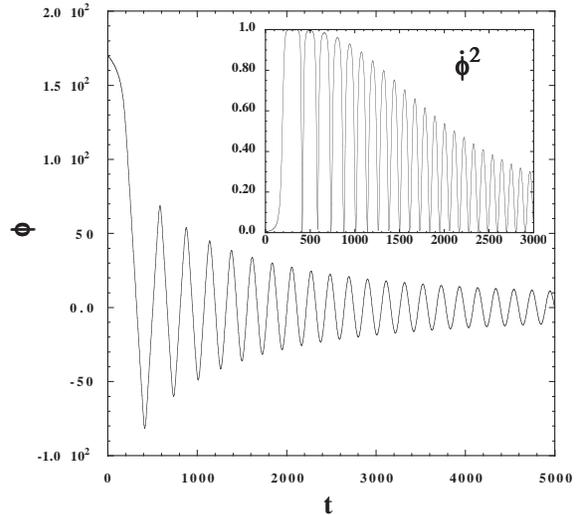}}
\caption{
The evolution of the Born-Infeld scalar field $\phi$
for $\beta=10^{-3}$, $T_3=1$ and $\Lambda=V_0$.
We start integrating from the beginning of inflation
with initial values $\phi_i=170$ and $\dot{\phi}_i=0$.
In this case we get the $e$-folding $N \sim 73$
around the end of inflation ($t \sim 200$).
This slow-roll stage is followed by a reheating phase
corresponding to the oscillation of the field $\phi$.
{\bf Inset}: The evolution of $\dot{\phi}^2$.
This quantity rapidly approaches 1 during the transition
from the inflationary phase to the reheating phase.
} \label{phievo}
\end{figure}

\begin{figure}
\resizebox{3.2in}{!}{\includegraphics{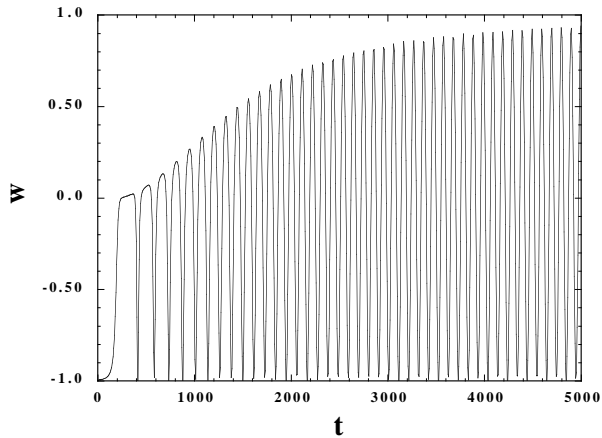}}
\caption{
The evolution of the equation of state parameter $w$
with same model parameters and initial conditions
as in Fig.~\ref{phievo}.
The equation of state is close to $-1$ during inflation,
which is followed by the oscillation of $\phi$
as the system enters the reheating stage.
At late times the average equation of state corresponds
to $\langle w \rangle =0$.
 } \label{omephi}
\end{figure}

\section{Late Time Evolution}

We have seen in the preceding sections that the massive scalar
field rolling on the D-brane can describe the inflation at early
epochs. We have shown that sufficient inflation can be drawn
assuming that the field begins rolling from larger values towards
the origin at which its potential has a minimum equal to
${\cal V}_{\rm dS}=\beta^2T_3-\Lambda$.

If the negative $\Lambda$ term is absent, the potential energy
at $\phi=0$ is ${\cal V}_{\rm dS}=\beta^2T_3$.
{}From the requirement of the amplitude of density perturbations
generated during inflation, we can determine the value once other
model parameters are fixed. For example, in the case of the
exponential potential (\ref{expo}), we found $\beta \sim 10^{-9}$
for the choice $\lambda=1$.
Then the potential energy at $\phi=0$
is $V_0 \sim 10^{-18} T_3$, which is much higher than the
critical density at present ($\rho_c \sim 10^{-47} {\rm GeV}^4$)
provided that $T_3$ is of order the Planck density. In this case,
although there exists a non-accelerating phase
with $\dot{\phi}^2>2/3$, the universe soon enters the regime of
an accelerated expansion as the field approaches the potential
minimum. Therefore we can not have a sufficient long period of
radiation and matter dominant stages.

When $\Lambda=0$ there exists another problem associated with
the equation of state for the field $\phi$.
{}From Eq.~(\ref{eqst}) one has $w=\dot{\phi}^2-1$
for $\Lambda=0$, which means that the equation of state
ranges $-1 \le w <0$ (see Fig.~\ref{bsw}). By using Eq.~(\ref{scaling}) we find
that the energy density of the field $\phi$ decreases slower than
$\rho \propto a^{-3}$.
Then this energy density
easily dominates the universe soon after inflation, thus disturbing
the thermal history of the universe.

This problem can be circumvented by implementing a negative
cosmological constant arising from the modulus stabilization.
We wish to consider a scenario in which
the present value of
the critical energy density, $\rho_c$,
can be reproduced when the field settles
down at the potential minimum ($\phi=0$), i.e.,
\begin{eqnarray}
\rho_c=\beta^2 T_3-\Lambda\,.
\end{eqnarray}
This corresponds to the
case where $\Lambda$ is very close to $\beta^2 T_3$.
Inclusion of this negative cosmological constant
drastically changes the dynamics of reheating.
We have $c=\Lambda/V \le 1$ during the reheating phase,
since $V$ is greater than $\beta^2 T_3$.
If $c$ is exactly one, one can show from
Eq.~(\ref{eqst}) that $w$ ranges $0 \le w \le 1$.
In this case the energy density of the field $\phi$ decreases
faster than $a^{-3}$ from Eq.~(\ref{scaling}).
More realistically $c$ is a function of time
satisfying $c \le 1$, in which case $w$ changes
between $-1$ and $1$.
Note that when $c$ is slightly less than $1$ the equation of state
reaches to $w=-1$ when the field crosses $\dot{\phi}^2=0$.

In Figs.~\ref{phievo} and \ref{omephi} we plot the evolution of
the field $\phi$ and the
equation of state $w$ for the potential
(\ref{eq03}) with $\beta=10^{-3}$,
$T_3=1$ and $\Lambda=V_0$.
As long as $\Lambda$ is very close to $V_0$, the
evolution of the system is very similar what is shown
in Figs.~\ref{phievo} and \ref{omephi}.
During the inflationary stage corresponding to $0 \le t
\lesssim 200$
in the figures, $\dot{\phi}^2$ is much smaller than unity.
This is followed by a reheating stage during which the field $\phi$
oscillates around the potential minimum.
As seen from the inset of Fig.~\ref{phievo}, $\dot{\phi}^2$ rapidly
grows toward 1 during the transition to the reheating phase.
{}From Fig.~\ref{omephi} we find that the equation of state
gradually enters the region with positive $w$
during reheating. At late stage $w$ oscillates between $-1$
and $1$, which yields the average equation of state
$\langle w \rangle=0$.
During the initial stage of reheating $\dot{\phi}^2$
is close to 1, which means that the equation of state
is approximately given as $w \approx -1+\dot{\phi}^2$
from Eq.~(\ref{eqst}). Therefore $w$ does not exceed 0
at the beginning of reheating, as seen in
Fig.~\ref{omephi}. However this picture changes with the
decrease of $\dot{\phi}^2$ and $w$ can take positive values.
When $\dot{\phi}^2$ becomes much smaller than 1, $w$ oscillates
between $-1$ and $1$, thus yielding the equation of state
for a pressure-less dust.
This is similar
to the standard reheating scenario with a massive
inflaton \cite{KLS}.

\begin{figure}
\resizebox{3.0in}{!}{\includegraphics{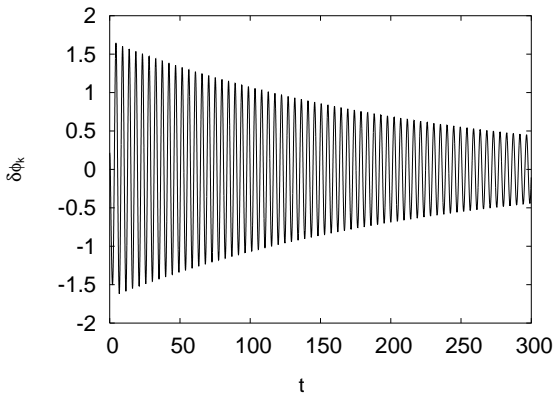}}
\caption{The evolution of perturbation $\delta{\phi}_k$ governed by Eq.~(\ref{phik}) for
the massive Born-Infeld scalar. Unlike the case of a tachyon field, the bahavor of perturbations
resembles with the evolution of  $\delta{\phi}_k$ for an ordinary massive scalar field}
\label{pert}
\end{figure}

Thus our scenario provides a satisfactory equation of state
during reheating unlike the case of the tachyon
inflation.
It was pointed out in Ref.~\cite{Frolov} that
there is a negative instability for
the tachyon fluctuations for the potential with a minimum, e.g.,
$V(\phi)=(1/2)m^2(\phi-\phi_0)^2$.
One may worry that this property may also hold in our scenario.
However this is not the case.
The each Fourier mode of the perturbation in $\phi$ satisfies
the following equation on the FRW background:
\begin{eqnarray}
\label{phik}
\frac{\ddot{\delta\phi_k}}{1-\dot{\phi}^2}
&+&
\left[3H+\frac{2\dot{\phi}\ddot{\phi}}{(1-\dot{\phi}^2)^2}
\right]\dot{\delta \phi_k} \nonumber \\
&+&\left[\frac{k^2}{a^2}+(\log
V),{}_{\phi\phi}\right]\delta\phi_k=0\,,
\end{eqnarray}
where $k$ is a comoving wavenumber.
This is the generalization of the perturbation equation
in Minkowski space time \cite{Frolov}.
While the $(\log V),{}_{\phi\phi}$ term is negative
for the potential $V(\phi)=(1/2)m^2(\phi-\phi_0)^2$,
our massive rolling scalar field potential (\ref{eq03})
corresponds to a positive mass with
$(\log V),{}_{\phi\phi}=\beta m^2$.
It is, therefore, not surprising that we do not have any instability for the
perturbation $\delta\phi_k$ in our case. The numerical treatment of Eq.~ (\ref{phik}) confirms
this behavior of field perturbations (see Fig.~\ref{pert}).
This is similar to what happens
in the case of the standard massive inflaton field,
thus showing a viability of our scenario.

The above good behavior originates from
the combination of the massive rolling scalar
potential and the negative cosmological constant
coming from the stabilization of the modulus.
It is also possible to explain the present critical
density $\rho_c$ when $\beta^2T_3$ is very close to
$\Lambda$.
Thus our scenario provides a sound mechanism
to explain inflation, reheating and dark energy
in the context of string theory.

\section{Conclusions and discussions}

In this paper we have presented a scenario based upon a massive
scalar field $\phi$ rolling on the D-brane which was shown in
Ref.~\cite{s2} as a possible solution in string theory. Using a
DBI type effective action presented in Ref.~\cite{mrg1}, we have
discussed the cosmological dynamics of the rolling scalar of the
anti-D3 brane of KKLT vacua, and have demonstrated that it can
lead to inflationary solutions at early epochs. In our scenario
it is possible to obtain a sufficient amount of inflation without
tuning a fundamental string scale, unlike the case of a rolling
tachyon \cite{KL}. The warp metric in the KKLT vacua provides us a
free parameter, the warp factor $\beta$. The presence of this
parameter allows us to obtain the COBE normalized value of density
perturbations without changing the brane tension and the masses
of massive string states.

We further investigated scalar and tensor perturbations generated
during the rolling scalar inflation and showed that our scenario
is compatible with recent observational data. It is interesting
to note that the steep inflation driven by an exponential
potential in the Randall-Sundrum braneworld II scenario is out of
the observational contour bounds \cite{cmbbrane}. This is
certainly related to the fact that the dynamics of the
Born-Infeld scalar is very different from that of an ordinary
scalar field as well as that of a scalar field on the brane.

We have also implemented the contribution of the negative
cosmological constant which arises from the stabilization of the
modulus fields \cite{KKLT}. Although this effect is not important
during inflation, the dynamics of reheating drastically changes
by taking into account the negative cosmological constant that
nearly cancels the potential energy of the massive rolling scalar
at the potential minimum. One of the problems of the tachyon
inflation is that the energy density of the tachyon scales slower
than that of the pressure less matter and the radiation
\cite{KL,Frolov}, which means that the tachyon over-dominates the
universe after inflation. In our scenario this problem is solved
by the negative cosmological constant. As shown in Sec.\,VII, the
average equation of state of the field $\phi$ approaches that of
the pressure-less dust during reheating. We also found that the
negative instabity of the field fluctuations in the tachyon case
\cite{Frolov} is not present for the  rolling massive scalar
field. This suggests that the reheating proceeds in a similar way
to the standard one driven by a massive inflaton field.

The massive rolling scalar can be used to explain the origin of
dark energy provided that the potential energy at the minimum
($\beta^2 T_3$) is very close to the amplitude of
the negative cosmological constant. The presence of the negative
cosmological constant is crucially important to lead to
a successful reheating and to explain the origin of
dark energy. Note that in the absence of the $-\Lambda$ term
it is difficult to obtain the amplitude of the
critical density unless we choose very small values of $\beta$.

While we performed a detailed analysis about the spectra of
scalar and tensor perturbations generated in the inflationary
stage, we did not precisely study the dynamics of reheating
including the decay of the field $\phi$. One interesting aspect
is the generation of gauge fields coupled to the rolling massive
scalar as was done in Ref.~\cite{jmc} in the tachyon case. We
leave the future work about the precise analysis of the reheating
dynamics in our scenario.

\section*{ACKNOWLEDGMENTS}
We thank Sam Leach, Antony Lewis, David Parkinson and Rita Sinha
for supporting the likelihood analysis. We are indebted to  R. Kallosh and A. Linde
for drawing our attention to the important problem of decompactification during inflation in the scenario
discussed in the present paper. The research of S.T. is
financially supported from JSPS (No.\,04942). S.T. thanks to all
members in IUCAA for their warm hospitality during which this
work was initiated. M.R.G. would like to thank A. Sen and S.
Parvizi for  useful discussion. MS acknowledges the helpful discussion with J. Maharana, T.
Padmanabhan, V. Sahni , N. Dadhich and T. Qureshi.


\begin{thebibliography}{99}

\bibitem{review}
{A.~Linde, {\em Particle Physics and Inflationary Cosmology},
Harwood, Chur (1990); A.~R.~Liddle and D.~H. ~Lyth, {\em
Cosmological inflation and large-scale structure}, Cambridge
University Press (2000).}

\bibitem{WMAP1}
{C.~L.~Bennett {\it et al.},
Astrophys.\ J.\ Suppl.\  {\bf 148}, 1 (2003)
[arXiv:astro-ph/0302207].}

\bibitem{WMAP2}
{D.~N.~Spergel {\it et al.},
Astrophys.\ J.\ Suppl.\  {\bf 148}, 175 (2003)
[arXiv:astro-ph/0302209].}

\bibitem{LR}
D.~H.~Lyth and A.~Riotto,
Phys.\ Rept.\  {\bf 314}, 1 (1999) [arXiv:hep-ph/9807278].


\bibitem{phiindustry}
V.~Sahni and A.~A.~Starobinsky,
Int.\ J.\ Mod.\ Phys.\ D {\bf 9}, 373 (2000)
[arXiv:astro-ph/9904398];T.~Padmanabhan,
Phys.\ Rept.\  {\bf 380}, 235 (2003) [arXiv:hep-th/0212290];
B.~Ratra and P.~J.~E.~Peebles,
Phys.\ Rev.\ D {\bf 37}, 3406 (1988); C.~Wetterich,
Nucl.\ Phys.\ B {\bf 302}, 668 (1988); J.~A.~Frieman, C.~T.~Hill,
A.~Stebbins and I.~Waga,
Phys.\ Rev.\ Lett.\  {\bf 75}, 2077 (1995)
[arXiv:astro-ph/9505060]; P.~G.~Ferreira and M.~Joyce,
Phys.\ Rev.\ Lett.\  {\bf 79}, 4740 (1997)
[arXiv:astro-ph/9707286]; I.~Zlatev, L.~M.~Wang and
P.~J.~Steinhardt,
Phys.\ Rev.\ Lett.\  {\bf 82}, 896 (1999)
[arXiv:astro-ph/9807002]; P.~Brax and J.~Martin,
Phys.\ Rev.\ D {\bf 61}, 103502 (2000) [arXiv:astro-ph/9912046];
T.~Barreiro, E.~J.~Copeland and N.~J.~Nunes,
Phys.\ Rev.\ D {\bf 61}, 127301 (2000) [arXiv:astro-ph/9910214];
A.~Albrecht and C.~Skordis,
Phys.\ Rev.\ Lett.\  {\bf 84}, 2076 (2000)
[arXiv:astro-ph/9908085]; V. Sahni, astro-ph/0403324; D.F. Mota, C. van de Bruck, astro-ph/0401504.


\bibitem{unifiedmodels}
P.~J.~Peebles and A.~Vilenkin, Phys.\ Rev.\ D {\bf 59}, 063505
(1999). E.J. Copeland, A. R. Liddle and J. E. Lidsey, Phys. Rev.
D {\bf 64}, 023509 (2001) [astro-ph/0006421]; G. Huey and Lidsey,
Phys. Lett. B {\bf 514},217(2001); V. Sahni, M. Sami and T.
Souradeep, Phys. Rev. D{\bf 65},023518(2002); A. S. Majumdar,
Phys.Rev. D{\bf 64} (2001) 083503; M. Sami, N. Dadhich and
Tetsuya Shiromizu, Phys.Lett. B{\bf 568}(2003)
118[hep-th/0304187]; M. Sami and V. Sahni, hep-th/0402086.
\bibitem{s1}
A. Sen, JHEP {\bf 0204}, 048 (2002)  [arXiv: hep-th/0203211]; JHEP
{\bf 0207}, 065 (2002) [ arXiv: hep-th/0203265]; Mod. Phys. Lett.
A {\bf 17}, 1797 (2002) [ arXiv: hep-th/0204143]; arXiv:
hep-th/0312153.

\bibitem{as1} A. Sen, JHEP {\bf 9910}, 008 (1999) [arXiv: hep-th/9909062];
M. R. Garousi, Nucl. Phys. B{\bf 584}, 284 (2000) [arXiv:
hep-th/0003122]; Nucl. Phys. B {\bf 647}, 117 (2002) [arXiv:
hep-th/0209068]; JHEP {\bf 0305}, 058 (2003) [arXiv:
hep-th/0304145]; E.A. Bergshoeff, M. de Roo, T.C. de Wit, E.
Eyras, S. Panda, JHEP {\bf 0005}, 009 (2000) [arXiv:
hep-th/0003221]; J. Kluson, Phys. Rev. D {\bf 62}, 126003 (2000)
[arXiv: hep-th/0004106]; D. Kutasov and V. Niarchos, Nucl. Phys.
B {\bf 666}, 56 (2003) [arXiv: hep-th/0304045].



\bibitem{tachyonindustry}
G.~W.~Gibbons,
Phys.\ Lett.\ B {\bf 537}, 1 (2002) [arXiv:hep-th/0204008];
M.~Fairbairn and M.~H.~G.~Tytgat,
Phys.\ Lett.\ B {\bf 546}, 1 (2002) [arXiv:hep-th/0204070];
S.~Mukohyama,
Phys.\ Rev.\ D {\bf 66}, 024009 (2002) [arXiv:hep-th/0204084];
A.~Feinstein,
Phys.\ Rev.\ D {\bf 66}, 063511 (2002) [arXiv:hep-th/0204140]; T.~Padmanabhan,
Phys.\ Rev.\ D {\bf 66}, 021301 (2002) [arXiv:hep-th/0204150]; D.~Choudhury,
D.~Ghoshal, D.~P.~Jatkar and S.~Panda,
Phys.\ Lett.\ B {\bf 544}, 231 (2002) [arXiv:hep-th/0204204];
G.~Shiu and I.~Wasserman,
Phys.\ Lett.\ B {\bf 541}, 6 (2002) [arXiv:hep-th/0205003];
T.~Padmanabhan and T.~R.~Choudhury,
Phys.\ Rev.\ D {\bf 66}, 081301 (2002) [arXiv:hep-th/0205055];
M.~Sami,
Mod.\ Phys.\ Lett.\ A {\bf 18}, 691 (2003) [arXiv:hep-th/0205146]; M.~Sami, P.~Chingangbam and T.~Qureshi,
Phys.\ Rev.\ D {\bf 66}, 043530 (2002) [arXiv:hep-th/0205179].
A.~Ishida and S.~Uehara,
JHEP {\bf 0302}, 050 (2003) [arXiv:hep-th/0301179];
JHEP {\bf 0210}, 034 (2002) [arXiv:hep-th/0207107]; A.~Mazumdar,
S.~Panda and A.~Perez-Lorenzana,
Nucl.\ Phys.\ B {\bf 614}, 101 (2001) [arXiv:hep-ph/0107058];
Y.~S.~Piao, R.~G.~Cai, X.~m.~Zhang and Y.~Z.~Zhang,
Phys.\ Rev.\ D {\bf 66}, 121301 (2002) [arXiv:hep-ph/0207143];
Z.~K.~Guo, Y.~S.~Piao, R.~G.~Cai and Y.~Z.~Zhang,
Phys.\ Rev.\ D {\bf 68}, 043508 (2003) [arXiv:hep-ph/0304236];
 G.~N.~Felder,
L.~Kofman and A.~Starobinsky,
JHEP {\bf 0209}, 026 (2002) [arXiv:hep-th/0208019]; B.~Wang,
E.~Abdalla and R.~K.~Su,
Mod.\ Phys.\ Lett.\ A {\bf 18}, 31 (2003) [arXiv:hep-th/0208023];
S.~Mukohyama,
Phys.\ Rev.\ D {\bf 66}, 123512 (2002) [arXiv:hep-th/0208094];
J.~g.~Hao and X.~z.~Li,
Phys.\ Rev.\ D {\bf 66}, 087301 (2002) [arXiv:hep-th/0209041];
M.~C.~Bento, O.~Bertolami and A.~A.~Sen,
Phys.\ Rev.\ D {\bf 67}, 063511 (2003) [arXiv:hep-th/0208124];
A.~Sen,
Int.\ J.\ Mod.\ Phys.\ A {\bf 18}, 4869 (2003)
[arXiv:hep-th/0209122]; C.~j.~Kim, H.~B.~Kim and Y.~b.~Kim,
Phys.\ Lett.\ B {\bf 552}, 111 (2003) [arXiv:hep-th/0210101];
J.~S.~Bagla, H.~K.~Jassal and T.~Padmanabhan,
Phys.\ Rev.\ D {\bf 67}, 063504 (2003) [arXiv:astro-ph/0212198];
F.~Leblond and A.~W.~Peet,
JHEP {\bf 0304}, 048 (2003) [arXiv:hep-th/0303035]; T.~Matsuda,
Phys.\ Rev.\ D {\bf 67}, 083519 (2003) [arXiv:hep-ph/0302035];
A.~Das and A.~DeBenedictis,
arXiv:gr-qc/0304017; A. DeBenedictis and A. Das,  gr-qc/0207077 ; M.~Majumdar and A.~C.~Davis,
arXiv:hep-th/0304226; S.~Nojiri and S.~D.~Odintsov,
Phys.\ Lett.\ B {\bf 571}, 1 (2003) [arXiv:hep-th/0306212];
D.~Bazeia, F.~A.~Brito and J.~R.~S.~Nascimento,
Phys.\ Rev.\ D {\bf 68}, 085007 (2003) [arXiv:hep-th/0306284];
L.~R.~W.~Abramo and F.~Finelli,
Phys.\ Lett.\ B {\bf 575}, 165 (2003) [arXiv:astro-ph/0307208]; L. Raul Abramo, Fabio Finelli, Thiago S. Pereira, astro-ph/0405041;
V.~Gorini, A.~Y.~Kamenshchik, U.~Moschella and V.~Pasquier,
arXiv:hep-th/0311111; M.~B.~Causse,
arXiv:astro-ph/0312206; P.~K.~Suresh,
arXiv:gr-qc/0309043; B.~C.~Paul and M.~Sami,
arXiv:hep-th/0312081; Jian-gang Hao, Xin-zhou Li, hep-th/0303093;
Xin-zhou Li, Dao-jun Liu, Jian-gang Hao,
Dao-jun Liu, Xin-zhou Li,
hep-th/0207146;  Zong-Kuan Guo, Yuan-Zhong Zhang, hep-th/0403151;
 Xin-zhou Li, Jian-gang Hao, Dao-jun Liu,  hep-th/0204252 ;  Dao-jun Liu, Xin-zhou Li,
astro-ph/0402063; J.M. Aguirregabiria, Ruth Lazkoz,  hep-th/0402190;  J.M. Aguirregabiria, Ruth Lazkoz, gr-qc/0402060 ;  Gianluca Calcagni, hep-ph/0402126; Kenji Hotta, hep-th/0403078 ; Xin-He Meng, Peng Wang, hep-ph/0312113.

\bibitem{KL}
L.~Kofman and A.~Linde, JHEP {\bf 0207}, 004 (2002)
[hep-th/0205121].

\bibitem{Frolov}
A.~V.~Frolov, L.~Kofman and A.~A.~Starobinsky,
Phys.\ Lett.\ B {\bf 545}, 8 (2002) [arXiv:hep-th/0204187].

\bibitem{s2}A. Sen, JHEP {\bf 0210}, 003 (2002)  [arXiv:hep-th/0207105].
\bibitem{jp}
J. Polchinski and Y. Cai, Nucl. Phys. B {\bf 296} (1988) 91; C.
G. Callan, C. Lovelace, C. R. Nappi and S. A. Yost, Nucl. Phys. B
{\bf 308} (1988) 221; M. Li, Nucl. Phys. B {\bf 460} (1996) 351
[arXiv:hep-th/9510161]; O. Bergman and M. R. Gaberdiel, Nucl.
Phys. B {\bf 499} 183 [arXiv:hep-th/9701137].

\bibitem{mbg}
M. B. Green and M. Gutperle, Nucl. Phys. B {\bf 476} (1996) 484
[arXiv:hep-th/9604091]; P. Di Vecchia, M. Frau, I. Pesando, A.
Lerda and R. Russo, Nucl. Phys. B {\bf 507} (1997) 259
[arXiv:hep-th/9707069]; P. Di Vecchia and A. Liccardo,
arXiv:hep-th/9912275.

\bibitem
{sen2} A. Sen, JHEP {\bf 0207} (2002) 065 [arXiv:hep-th/0203265].


\bibitem{comment}
In spatially homogeneous field configuration, a solution is
characterized by initial position and velocity of the scalar
field $\phi$. We choose the origin of time when velocity of the
scalar field is zero. Hence, the homogeneous solution is
characterized by one parameter, the initial position of  scalar
field.

\bibitem{witten}E. Witten, Nucl. Phys. B {\bf 460}, 335 (1996)
[arXiv: hep-th/9510135].
\bibitem{nsw}N. Seiberg and E. Witten, JHEP {\bf 9909}, 032 (1999)
[arXiv: hep-th/9908142].
\bibitem{mrg1}
M. R. Garousi, JHEP {\bf 0312} (2003) 035 [arXiv:hep-th/0307197].

\bibitem{mrgrm}M. R. Garousi and R. C. Myers, Nucl. Phys. B {\bf
475}, 193 (1996) [arXiv: hep-th/9603194]; A. Hashimoto and I. R.
Klebanov, Phys. Lett. B {\bf 381}, 437 (1996) [arXiv:
hep-th/9604065]; Nucl. Phys.  (Proc. Suppl.) B {\bf 55}, 118
(1997) [arXiv: hep-th/9611214].

\bibitem{mrg2}M. R. Garousi, JHEP {\bf 9812}, 008 (1998) [arXiv:
hep-th/980578]; Nucl. Phys. B {\bf 579}, 209 (2000) [arXiv:
hep-th/9909214].

\bibitem{bsa}B.S. Acharya, [arXiv: hep-th/0212294];
S. Hellerman, J. McGreevy and B. Williams, [arXiv:
hep-th/0208174]; A. Dabholkar and C. Hull, [arXiv:
hep-th/0210209].
\bibitem{KKLT} S. Kachru, R. Kallosh, A. Linde
and S. P. Trivedi, Phys. Rev. D {\bf 68}, 046005 (2003) [arXiv:
hep-th/0301240].


\bibitem{KKLMMT} S. Kachru, R. Kallosh, A. Linde,
J. Maldacena, L. McAllister and S. P. Trivedi, JCAP {\bf 0310},
013 (2003) [arXiv: hep-th/0308055].
\bibitem{Linde} R. Kallosh and A. Linde, Private communication: Throughout the paper we have assumed that the volume of internal space is
frozen as in KKLT model. When inflaton field is around zero, it was shown in
[23] that the non-perturbative correction (17) to the tree level
superpotential makes it possible to have a positive minimum at finite
volume. This is in part due to the fact that the anti-D3 brane correction
(19) is small. During the inflation the anti-D3 brane correction is again
like (19) in which $T_3$ should be replaced by
$T_3\exp(a_0^2m^2\phi^2/(\sqrt{2}g_s^2(Im\rho)^{3/2}))$. With this
correction, however, one may not get anymore a local minimum  for the
supergravity potential during inflation in which  $\phi$ is very large. In
this case one may need to consider also $\alpha'$ corrections to the Kahler
potential \cite{becker} to study stabilization of the volume of internal space.
In our opinion, the problem of volume stabilization in the
model presented in this paper, is important and requires further investigation.
\bibitem{becker} K. Becker, M. Becker, M. Haack and J. Louis, JHEP  {\bf 0206}, 060 (2002),
[arXiv: hep-th/0204254].

\bibitem{pm}K. Dasgupta, G. Rajesh and Sethi, JHEP {\bf 0008}, 023 (1999)
[arXiv: hep-th/9908088]; K. Becker and M. Becker, Nucl. Phys. B
{\bf 477}, 155 (1996) [arXiv: hep-th/9605053]; H. Verlinde, Nucl.
Phys. B {\bf 580}, 264 (2000) [arXiv: hep-th/9906182]; C. Chan,
P. Paul and H. Verlinde, Nucl. Phys. B {\bf 581}, 156 (2000)
[arXiv: hep-th/0003236]; P. Mayr, Nucl. Phys. B {\bf 593}, 99
(2001) [arXiv: hep-th/0003198]; JHEP {\bf 0011}, 013 (2000)
[arXiv: hep-th/0006204]; B. Greene, K. Schalm and G. Shiu, Nucl.
Phys. B {\bf 584}, 480 (2000) [arXiv: hep-th/0004103].

\bibitem{sg}S. Giddings, S. Kachru and J. Polchinski,
Phys. Rev. D {\bf 66}, 106006 (2002) [arXiv: hep-th/0105079].

\bibitem{ew}
E. Witten, Nucl. Phys. B {\bf 474}, 343 (1996) [arXiv:
hep-th/9604030].

\bibitem{sk}
S. Kachru, J. Pearson and H. Verlinde, JHEP {\bf 0206}, 021
(2002) [arXiv: hep-th/0112197].

\bibitem{jmc}
G. Shiu, S.-H. H. Tye and I. Wasserman, Phys. Rev. D {\bf 67},
083517 (2003) [arXiv:hep-th/0207119]; J.M. Cline, H. Firouzjahi
and P. Martineau, JHEP {\bf 12}, 999 (2001) [arXiv:
hep-th/0207156]; N. Barnaby and J.M. Cline, arXiv: hep-th/0403223.

\bibitem{Hwang}
J.~c.~Hwang and H.~Noh,
Phys.\ Rev.\ D {\bf 66}, 084009 (2002)
[arXiv:hep-th/0206100].


\bibitem{MFB}
V.~F.~Mukhanov, H.~A.~Feldman and R.~H.~Brandenberger,
Phys.\ Rept.\  {\bf 215}, 203 (1992).

\bibitem{comment2}
Note that in some other cosmological models $c_{\rm s}^2$
can be negative. See, e.g., Ref.\,\cite{Shinji}.

\bibitem{Shinji}
S.~Tsujikawa, R.~Brandenberger and F.~Finelli,
Phys.\ Rev.\ D {\bf 66}, 083513 (2002)
[arXiv:hep-th/0207228].


\bibitem{cmb1}
H.~V.~Peiris {\it et al.},
Astrophys.\ J.\ Suppl.\  {\bf 148}, 213 (2003)
[arXiv:astro-ph/0302225];
V.~Barger, H.~S.~Lee, and D.~Marfatia,
Phys.\ Lett.\ B {\bf 565}, 33 (2003)
[arXiv:hep-ph/0302150];
M.~Tegmark {\it et al.}  [SDSS Collaboration],
arXiv:astro-ph/0310723.

\bibitem{cmb2}
W.~H.~Kinney, E.~W.~Kolb,
A.~Melchiorri, and A.~Riotto,
arXiv:hep-ph/0305130;
S.~M.~Leach and A.~R.~Liddle,
arXiv:astro-ph/0306305.

\bibitem{cmbbrane}
A.~R.~Liddle and A.~J.~Smith,
Phys.\ Rev.\ D {\bf 68}, 061301 (2003)
[arXiv:astro-ph/0307017];
S.~Tsujikawa and A.~R.~Liddle,
arXiv:astro-ph/0312162.

\bibitem{SV}
D.~A.~Steer and F.~Vernizzi,
arXiv:hep-th/0310139.

\bibitem{Burin}
S.~Tsujikawa and B.~Gumjudpai,
arXiv:astro-ph/0402185, Physical Review D to appear.

\bibitem{KLS}
J.~H.~Traschen and R.~H.~Brandenberger,
Phys.\ Rev.\ D {\bf 42}, 2491 (1990);
Y.~Shtanov, J.~H.~Traschen and R.~H.~Brandenberger,
Phys.\ Rev.\ D {\bf 51}, 5438 (1995);
L.~Kofman, A.~D.~Linde and A.~A.~Starobinsky,
Phys.\ Rev.\ Lett.\  {\bf 73}, 3195 (1994);
Phys.\ Rev.\ D {\bf 56}, 3258 (1997).



\end{thebibliography}
\end{document}